\documentclass{amsart}
\usepackage{graphicx}
\newtheorem{thm}{Theorem}[section]

\newtheorem{prop}[thm]{Proposition}
\theoremstyle{definition}

\theoremstyle{remark}
\newtheorem{rem}[thm]{Remark}
\numberwithin{equation}{section}

\newcommand{\A}{\mathcal{A}}
\begin{document}

\title[]{Non-negative perturbations of non-negative self-adjoint operators}%
\author{Vadym Adamyan}%
\address{Odessa National I.I. Mechnikov University, Odessa 65026, Ukraine}%
\email{vadamyan@paco.net}%

\thanks{The author is grateful to Sergey Gredescul for the idea of this work and Yury Arlinskii for valuable discussion. The author acknowledges the support from USA Civil Research and Development Foundation(CRDF) and Government of Ukraine grant UM2-2811-OD06.}%
\subjclass{Primary 47A63, 47B25; Secondary 47B65}
\keywords{Non-negative self-adjoint extension, non-negative contraction, singular perturbation}
\begin{abstract}
Let $A$ be a non-negative self-adjoint operator in a Hilbert space $\mathcal{H}$ and $A_{0}$ be some densely defined closed restriction of $A_{0}$, $A_{0}\subseteq A \neq A_{0}$. It is of interest to know whether $A$ is the unique non-negative self-adjoint extensions of $A_{0}$ in $\mathcal{H}$. We give a natural criterion that this is the case and if it fails, we describe all non-negative extensions of $A_{0}$. The obtained results are applied to investigation of  non-negative singular point perturbations of the Laplace and poly-harmonic operators in $\mathbb{L}_{2}(\mathbf{R}_{n})$.
\end{abstract}
\maketitle
\section{Introduction}
In this paper we deal with a non-negative self-adjoint operator $A$ in a Hilbert space$\mathcal{H}$, some densely defined not essentially self-adjoint restriction $A_{0}$ of $A$ and again with self-adjoint extensions of $A_{0}$ in $\mathcal{H}$, which following \cite{AGHH} we call here \textit{singular perturbations} of $A$. For quick getting onto the matter of main problem let us compare the point perturbations of self-adjoint Laplace operators $-\Delta$ in three and two dimensions acting in $\mathbb{L}_{2}(\mathbf{R}_{3})$ and $\mathbb{L}_{2}(\mathbf{R}_{2})$, respectively, that is let us consider the restriction $-\Delta^{0}$ of $-\Delta$ onto the Sobolev subspaces $\mathbb{H}_{2}^{2}\left( \mathbf{R}_{i}\setminus\{0\}\right), \, i=3,2$ and self-adjoint extensions $-\Delta_{\alpha}, \, \alpha\in \mathbf{R}$  of $-\Delta^{0}$ in $\mathbb{L}_{2}(\mathbf{R}_{i})$ with domains
\begin{equation} \label{dom}
\begin{array}{c}
\mathcal{D}_{\alpha}^{(3)}:=\left\{f: \, f\in \mathbb{H}_{2}^{2}\left(\mathbf{R}_{3}\right), \, \underset
{|\mathbf{x}|\downarrow 0}{\lim} \left[\frac {d} {d|\mathbf{x}|}\left(|\mathbf{x}|f(\mathbf{x})\right)-\alpha |\mathbf{x}|f(\mathbf{x})\right]=0 \right\}, \\
\mathcal{D}_{\alpha}^{(2)}:=\left\{f: \, f\in \mathbb{H}_{2}^{2}\left(\mathbf{R}_{2}\right), \, \underset
{|\mathbf{x}|\downarrow 0}{\lim}\left[\left(\frac{2\pi\alpha}{\ln|\mathbf{x}|}+1\right)f(\mathbf{x})- \underset
{|\mathbf{x'}|\downarrow 0}{\lim}\frac{\ln|\mathbf{x}|}{\ln|\mathbf{x'}|}f(\mathbf{x'})\right]=0 \right\}.
\end{array}
\end{equation}
The self-adjoint operators $-\Delta_{\alpha}$ are just mentioned above singular perturbations of $-\Delta$. Resolvents $\left(-\Delta_{\alpha}-z\right)^{-1}, \, z\in\rho(-\Delta_{\alpha}),$ of operators $-\Delta_{\alpha}$ act in the corresponding spaces  $\mathbb{L}_{2}$ as integral operators with kernels (Green functions)  \cite{AGHH}:
\begin{equation}\label{Green3}
G_{\alpha,z}^{3}(\mathbf{x},\mathbf{x'})=\left\{\begin{array}{c}
G_{z}^{(0)}(\mathbf{x},\mathbf{x'}) +(\alpha-i\sqrt{z}/4\pi)^{-1}G_{z}^{(0)}(\mathbf{x},0)G_{z}^{(0)}(0,\mathbf{x'}),\\
G_{z}^{(0)}(\mathbf{x},\mathbf{x'})=\frac{\exp{i\sqrt{z}|\mathbf{x}-\mathbf{x'}|}}{4\pi|\mathbf{x}-\mathbf{x'}|} \:(\mathrm{three\, dimension});
\end{array} \right.
\end{equation}
\begin{equation}\label{Green2}
G_{\alpha,z}^{2}(\mathbf{x},\mathbf{x'})=\left\{ \begin{array}{c}
G_{z}^{(0)}(\mathbf{x},\mathbf{x'}) +
2\pi(2\pi\alpha-\psi (1) +\ln{\sqrt{z}/2i})^{-1}
G_{z}^{(0)}(\mathbf{x},0)G_{z}^{(0)}(0,\mathbf{x'}),\\
G_{z}^{(0)}(\mathbf{x},\mathbf{x'})=(\frac{i}{4})H_{0}^{(1)}(i\sqrt{z}|\mathbf{x}-\mathbf{x'}|) \:(\mathrm{two\, dimension}).
\end{array} \right .
\end{equation}
By (\ref{Green3}) the Green function $G_{\alpha,z}(\mathbf{x},\mathbf{x'})$ of self-adjoint operator $-\Delta_{\alpha}$ in $\mathbb{L}_{2}(\mathbf{R}_{3})$) is holomorphic on the half-axis $(-\infty,0)$ for $\alpha \geq 0$ and has on this half-axis a simple pole for $\alpha <0$. Hence in the case of three dimensions self-adjoint extensions $-\Delta_{\alpha}$ are non-negative for all $(\alpha \geq 0)$ and non-positive for $\alpha\ < 0$.

Contrary to this by (\ref{Green2}) in the case of two dimensions for any $\alpha \in \mathbf{R}$ the Green function $G_{\alpha,z}$ has a simple pole on the half-axis $(-\infty,0)$/. Hence all singular perturbations $-\Delta _{\alpha}$ of the two-dimensional Laplace operators have one negative eigenvalue. In other words the standardly defined Laplace operator $-\Delta$ is the unique non-negative self-adjoint extension in $\mathbb{L}_{2}(\mathbf{R}_{2})$ of the symmetric operator $-\Delta^{0}$ in $\mathbb{L}_{2}(\mathbf{R}_{2})$.\footnote{The attention of author to this phenomenon was drawn by Sergey Gredeskul.}

In this note we try to reveal the underlying cause of such discrepancy. Remind that each densely defined non-negative symmetric operator has at least one non-negative canonical self-adjoint extension (Friedrichs extension). In more general setting we try to understand here why in some cases the non-negative extension appears to be unique. Actually this questions is embedded into the framework of the general extension theory for semi-bounded symmetric operators developed in the famous paper of M.G. Krein \cite{K}. Naturally, there is a criterium of uniqueness of non-negative extension in \cite{K}. In the next Section using only approaches of \cite{K} we find another form of this criterium directly facilitated to investigation of singular perturbations and for cases where conditions of these criterium fail describe all non-negative singular perturbations of a given non-negative self-adjoint operator $A$ associated with some its densely defined non-self-adjoint restriction $A_{0}$. In fact we give here a parametrization of the operator interval $[A_{\mu},A_{M}]$ of all canonical non-negative self-adjoint extensions of a given densely defined non-negative operator. The third Section illustrates obtained results by  the example of singular perturbations of  Laplace and poly-harmonic operators in $\mathbb{L}_{2}(\mathbf{R}_{n})$.

Note that very close results were obtained recently in somewhat different way in \cite{ATs}, where in terms of this note were described singular perturbations of the Friedrichs extension of a given densely defined non-negative operator and also with illustration by the example of singular perturbations of the Laplace operator in $\mathbb{L}_{2}(\mathbf{R}_{3})$.
\section{Uniqueness criterium and parametrization of non-negative singular perturbations }
Let $A$ be a non-negative self-adjoint operator  acting in the Hilbert space $\mathcal{H}$ and $A_{0}$ be a densely defined closed operator, which is a restriction of $A$ onto a subset $\mathcal{D}(A_{0})$ of the domain $\mathcal{D}(A)$ of $A$. Let us consider the subspaces $\mathcal{M}:=(I+A_{0}\mathcal{D}(A_{0})$ and $\mathcal{N}:=\mathcal{H}\ominus \mathcal{M}$. We will assume that
\begin{equation}\label{propert}
\begin{array}{ccc}
  1) \; \mathcal{M}\neq \mathcal{H}, & 2) \; \mathcal{N}\cap\mathcal{D}(A)=\{0\}.
\end{array}
\end{equation}
 We call all self-adjoint extensions of $A_{0}$ in $\mathcal{H}$ other than the given $A$ singular perturbations of $A$. It is of interest to know whether there are non-negative operators among singular perturbations of $A$. In this section we try to find a convenient criterium that such singular perturbations of $A$ does not exist. In other words we look for a criterium that $A$ is one and only non-negative operator among all self-adjoint extensions of $A_{0}$.
Following the approach developed in the renowned paper of M.G. Krein \cite{K} let us consider the operator from $K_{0}:\mathcal{M} \rightarrow \mathcal{H}$ defined by relations
\begin{equation} \label{contr}\begin{array}{ccc}
 f=\left(I+A_{0}\right)x, & K_{0}f=\A_{0}x, & x\in \mathcal{D}(A_{0}).
 \end{array}
 \end{equation}
 It is easy to see that $K_{0}$ is a non-negative contraction:
 \begin{equation} \label{contr1} \begin{array}{ccc}
 \left(K_{0}f,f\right)\geq 0, & \|K_{0}f\|^{2}\leq \|f\|^{2}, & f \in \mathcal{M}.
 \end{array}
 \end {equation}
 Let $A_{1}$ be any non-negative self-adjoint extension of $A_{0}$ in $\mathcal{H}$. Then $K_{1}:=A_{1}\left(A_{1}+I\right)^{-1}$ is a non-negative operator, which is a contactive extension of $K_{0}$ from the domain $\mathcal{M}$ onto the whole $\mathcal{H}$, $K_{1}f=K_{0}f, \, f\in \mathcal{M}$.

 From the other hand for any contractive extension $K_{1}$ from $\mathcal{M}$ onto $\mathcal{H}$ such that the unity is not its eigenvalue the non-negative self-adjoint operator $A_{1}=K_{1}\left(I-K_{1}\right)^{-1}$ is a self-adjoint extension of $A_{0}$ in $\mathcal{H}$. Therefore  \textit{$A_{0}$ has unique non-negative self-adjoint extension in $\mathcal{H}$ if and only if $K_{0}$ admits only one non-negative contractive extension onto the whole $\mathcal{H}$, no eigenvalue of which $=1$, that is $K=A(I+A)^{-1}$}.
 So the uniqueness of $A$ as non-negative extension of $A_{0}$ is equivalent to uniqueness of $K_{0}$ as non-negative contractive extension of $K_{0}$.

 From now on we will denote by $\mathbf{G}$ the set consisting of $A$ and all its singular perturbations and by $\mathbf{C}$ the set of non-negative contractions obtained from $\mathbf{G}$ by transformation $A_{1}\rightarrow A_{1}\left(A_{1}+I\right)^{-1}, \, A_{1}\in \mathbf{G}$. Let us denote by $P_{\mathcal{M}}$ the orthogonal projector onto $\mathcal{M}$ in $\mathcal{H}$ and let $P_{\mathcal{N}}=I-P_{\mathcal{M}}$. With respect to representation of $\mathcal{H}$ as the orthogonal sum $\mathcal{M}\oplus\mathcal{N}$ we can represent each operator from $ \mathbf{C}$ as $2\times 2$ block operator matrix
 \begin{equation} \label{block}
 K_{X}=\left(
   \begin{array}{cc}
     T & \Gamma^{*} \\
     \Gamma & X \\
   \end{array}
 \right)
 \end{equation}
 Here
 \[ \begin{array}{cc}
      T=P_{\mathcal{M}}K_{0}|_{\mathcal{M}}, & \Gamma=P_{\mathcal{M}}K_{0}|_{\mathcal{M}}.
    \end{array} \]
and $X$ is some non-negative contraction in $\mathcal{N}$, which distinguishes different elements from $\mathbf{C}$. Since each $K_{X}\in\mathbf{C}$
is non-negative and contractive then
\begin{equation}\label{necess} \begin{array}{cc}
                 T\geq 0 ;& T^{2}+\Gamma ^{*}\Gamma  \leq I
               \end{array}
\end{equation}
Note further that

$K_{X}\in \mathbf{C}$ \textit{ is equivalent to
\begin{equation}\label{triv}
                               \begin{array}{cc}
                 K_{X}+\varepsilon I \geq 0  ;& (1+\varepsilon)I-K_{X}\geq 0 \;
               \end{array}
\end{equation}
for any } $ \varepsilon >0$.

The block matrix representation  of $K_{X}$ and the Schur -Frobenius factorization formula transform (\ref{triv}) into the following block matrix inequalities:
\begin{equation} \label{ineq1}
\begin{array}{c}
  \left(
  \begin{array}{cc}
    I & 0 \\
    \Gamma (T+\varepsilon)^{-1} & I \\
  \end{array}
\right)
\left(
  \begin{array}{cc}
    T +\varepsilon& 0 \\
    0 & X+\varepsilon-\Gamma(T+\varepsilon)^{-1}\Gamma^{*} \\
  \end{array}
\right)\times \\
  \left(
  \begin{array}{cc}
    I & (T+\varepsilon)^{-1}\Gamma^{*} \\
    0 & I \\
  \end{array}
\right)\geq 0,
\end{array}
\end{equation}

\begin{equation} \label{ineq2}
\begin{array}{c}
  \left(
  \begin{array}{cc}
    I & 0 \\
    -\Gamma (I+\varepsilon-T)^{-1} & I \\
  \end{array}
\right)
\left(
  \begin{array}{cc}
     1+\varepsilon-T& 0 \\
    0 & 1+\varepsilon-X-\Gamma(1+\varepsilon-T)^{-1}\Gamma^{*} \\
  \end{array}
\right)\times \\
  \left(
  \begin{array}{cc}
    I & -(1+\varepsilon-T)^{-1}\Gamma^{*} \\
    0 & I \\
  \end{array}
\right)\geq 0.
\end{array}
\end{equation}
By our assumptions $T\geq 0$ and $I-T\geq 0$. Therefore block matrix inequalities (\ref{ineq1}) and (\ref{ineq2}) are reduced to
 \begin{equation}\label{ineq3}
 \left\{ \begin{array}{c}
   X+\varepsilon I-\Gamma (T+\varepsilon I)^{-1}\Gamma^{*}\geq 0, \\
   (1+\varepsilon )I-X-\Gamma[(1+\varepsilon )I-T]^{-1}\Gamma^{*}\geq 0, \; \varepsilon >0.
 \end{array}\right.
 \end{equation}
 Observe that operator functions of $\varepsilon$ in the left hand sides of inequalities (\ref{ineq3}) are monotone. Setting
 \[
 Y:=X-\underset{\varepsilon\downarrow 0}{\lim}\,\Gamma (T+\varepsilon I)^{-1}\Gamma^{*}
 \]
 we conclude from (\ref{ineq3}) that $K_{X} \in \mathbf{C}$ if and only if
 \begin{equation}\label{ineq4}
 0\leq Y \leq I-\underset{\varepsilon\downarrow 0}{\lim}\,\left\{ \Gamma (T+\varepsilon I)^{-1}\Gamma^{*}+\Gamma[(1+\varepsilon )I-T]^{-1}\Gamma^{*}\right\}.
 \end{equation}
  Hence the equality
  \begin{equation}\label{crit1}
  I-\underset{\varepsilon\downarrow 0}{\lim}\,\left\{ \Gamma (T+\varepsilon I)^{-1}\Gamma^{*}+\Gamma[(1+\varepsilon )I-T]^{-1}\Gamma^{*}\right\}=0
  \end{equation}
 is the criterium that there are no contractive non-negative extension of $K_{0}$ in $\mathcal{H}$ other than $K$.

 Let us express now (\ref{ineq4}) and {\ref{crit1}) in terms of given $K$ and $A$. To this end we use the following proposition.
 \begin{prop}
 Let $L$ be a bounded invertible operator in the Hilbert space $\mathcal{H}=\mathcal{M}\oplus\mathcal{N}$ given as $2\times@$ block operator matrix,
 \[
 L=\left(
     \begin{array}{cc}
       R & U \\
       V & S \\
     \end{array}
   \right)
 \], where $R$ and $S$ are invertible operators in $\mathcal{M}$ and $\mathcal{N}$, respectively, and $U,V$ act between $\mathcal{M}$ and $\mathcal{N}$. If $R$ is invertible operator in $\mathcal{M}$, then
 \begin{equation}\label{inverse}
 \left(
   \begin{array}{cc}
     R^{-1} & 0 \\
     0 & 0 \\
   \end{array}
 \right)=L^{-1}-L^{-1}P_{\mathcal{N}}\Lambda ^{-1}P_{\mathcal{N}}L^{-1},
 \; \Lambda=P_{\mathcal{N}}L^{-1}|_{\mathcal {N}}.
 \end{equation}
 \end{prop}
 Setting
 \begin{equation}\label{block2} \begin{array}{cc}
 \Lambda_{1,\varepsilon}=P_{\mathcal{N}}(K+\varepsilon I)^{-1}|_{\mathcal{N}} & \Lambda_{2,\varepsilon}=P_{\mathcal{N}}[(1+\varepsilon)I-K]^{-1}|_{\mathcal{N}}
  \end{array}
 \end{equation}
 and applying (\ref{inverse}) with $L=K+\varepsilon I$ and
 \[ \begin{array}{cc}
     R=T+\varepsilon I & U=\Gamma^{*}=P_{\mathcal{M}}K|_{\mathcal{N}}= P_{\mathcal{M}}[K+\varepsilon I]|_{\mathcal{N}}\\
    V=\Gamma=P_{\mathcal{N}}K|_{\mathcal{M}}=P_{\mathcal{N}}[K+\varepsilon I]|_{\mathcal{M}} & S=P_{\mathcal{N}}K|_{\mathcal{N}}+\varepsilon I
   \end{array}
 \]
 yields
 \[
 \Gamma (T+\varepsilon I)^{-1}\Gamma^{*}=P_{\mathcal{N}}K|_{\mathcal{N}}+\varepsilon I-\Lambda_{1,\varepsilon}^{-1}.
 \]
 In the same fashion we get
 \[
 \Gamma [(1+\varepsilon)I-T]^{-1}\Gamma^{*}=P_{\mathcal{N}}[I-K]|_{\mathcal{N}}+\varepsilon I-\Lambda_{2,\varepsilon}^{-1}.
 \]
 Hence
 \begin{equation}\label{block4}
 I-\underset{\varepsilon\downarrow 0}{\lim}\,\left( \Gamma (T+\varepsilon I)^{-1}\Gamma^{*}+\Gamma[(1+\varepsilon )I-T]^{-1}\Gamma^{*}\right)=\underset{\varepsilon\downarrow 0}{\lim}\,\Lambda_{1,\varepsilon}^{-1}+\underset{\varepsilon\downarrow 0}{\lim}\,\Lambda_{2,\varepsilon}^{-1}.
 \end{equation}
 Combining (\ref{ineq4}), (\ref{crit1}) and (\ref{block4}) results in the following theorem.
 \begin{thm} \label{contractive}
 Let $K$ be a non-negative contraction in the Hilbert space $\mathcal{H}=\mathcal{M}\oplus \mathcal{N}$, $K_{0}$ is the restriction of $K$ onto the subspace $\mathcal{M} (=\mathcal{M}\oplus \{0 \} )$ and
 \[\begin{array}{cc}
    G_{1}= \underset{\varepsilon\downarrow 0}{\lim}\left(P_{\mathcal{N}}[K+\varepsilon I]|_{\mathcal{N}}\right)^{-1} & G_{2}= \underset{\varepsilon\downarrow 0}{\lim}\left(P_{\mathcal{N}}[I-K+\varepsilon I]|_{\mathcal{N}}\right)^{-1}
   \end{array}
 \]
 Then the set $\mathbf{C}$ of all non-negative contractive extensions $K_{X}$ of $K_{0}$ in $\mathcal{H}$ is described by expression
 \begin{equation} \label{represent}
 K_{X}=\left(
   \begin{array}{cc}
     P_{\mathcal{M}}K|_{\mathcal{M}} & P_{\mathcal{M}}K|_{\mathcal{N}} \\
     P_{\mathcal{M}}K|_{\mathcal{N}} & X \\
   \end{array}
 \right),
 \end{equation}
 where $X$ runs the set of all non-negative contractions in $\mathcal{N}$ satisfying inequalities
 \begin{equation} \label{main}
 P_{\mathcal{N}}K|_{\mathcal{N}}-G_{1}\leq X \leq P_{\mathcal{N}}K|_{\mathcal{N}}+G_{2}.
 \end{equation}
 In particular, $K$ is the unique non-negative contractive extension of $K_{0}$ if and only if $G_{1}=G_{2}=0$.
 \end{thm}
 \begin{rem}
 The set  $\mathbf{C}$ of non-negative contractive of $K_{0}$ contains the minimal extension $K_{X_{\mu}}$ with $X_{\mu}=P_{\mathcal{N}}K|_{\mathcal{N}}-G_{1}$ in (\ref{represent}) and the maximal extension $K_{X_{M}}$ with $X_{M}=P_{\mathcal{N}}K|_{\mathcal{N}}+G_{2}$ in (\ref{represent}. If $G_{1}=0 \, (G_{2}=0)$, then $K$ is the minimal (maximal) element of $\mathbf{C}$.
 \end{rem}
 Theorem \ref{contractive} can be formulated in terms of non-negative self-adjoint operator $A$ and its non-negative singular perturbations.
 \begin{thm}\label{singular}
 Let $A$ be a non-negative self-adjoint operator in the Hilbert space $\mathcal{H}$, $A_{0}$ is a densely defined closed symmetric operator, which is a restriction of $A$ onto a linear subset $\mathcal{D}(A_{0})\subset \mathcal{D}(A)$ such that $\mathcal{N}=(I+A)\mathcal{D}(A_{0})\neq \{ 0\}$ and let
 \[\begin{array}{cc}
    G_{1}= \underset{\varepsilon\downarrow 0}{\lim}\left(P_{\mathcal{N}}[I+A][A+\varepsilon I]|_{\mathcal{N}}\right)^{-1} & G_{2}= \underset{\varepsilon\downarrow 0}{\lim}\left(P_{\mathcal{N}}[I+A][I+\varepsilon A]|_{\mathcal{N}}\right)^{-1}
   \end{array}
 \]
 Then the set of all non-negative singular perturbations $A_{Y}$ of $A$ is described by the formula
 \begin{equation} \label{descr}
 \left\{ \begin{array}{c}
   f=g-Y(I+A)g, \\
   A_{Y}f=Ag+Y(I+A)g,
 \end{array}\right.
 \end{equation}
 where $g\in \mathcal{D}(A)$ and $Y$ runs the set of non-negative contractions in $\mathcal{N}$ satisfying inequalities
  \begin{equation}
 -G_{1}\leq Y \leq G_{2}.
\end{equation}
 $A$ has no singular non-negative perturbations if and only if $G_{1}=G_{2}=0$.
 \end{thm}
 \begin{rem}\label{singular1}
 The set of all non-negative singular perturbations of $A$ contains the minimal perturbation $A_{\mu}$ with and the maximal perturbation $A_{M}$ such that any non-negative perturbation $A_{X}$ satisfies inequalities
 \[
 \left(I+A_{M}\right)^{-1}\leq A_{Y}\leq \left(I+A_{\mu}\right)^{-1}.
 \]
 The corresponding values of parameters $Y$ in Theorem \ref{singular} are
 \begin{equation}\label{extreme}
  \begin{array}{c}
                 Y_{\mu}=-G_{1} \\
              Y_{M}=G_{2}
            \end{array}
 \end{equation}
 If $G_{1}=0 \, (G_{2}=0)$, then the minimal (maximal) perturbation coincides with $A$.
 \end{rem}
 By simple calculation we get from (\ref{descr}) the following version of the M.G. Krein resolvent formula.
 \begin{prop}
 The set of resolvents of all non-negative singular perturbations $A_{Y}$ of $A$ is described by the M.G. Krein formula
 \begin{equation}\label{Krein}
 \begin{array}{c}
   \left(A_{Y}-zI\right)^{-1}=\left(A-zI\right)^{-1} \\
   -(1+z)(A+I)(A-zI)^{-1}Y\left[I+(1+z) P_{\mathcal{N}} (A+I)(A-zI)^{-1}Y \right]^{-1} \times \\
   P_{\mathcal{N}}(A+I)(A-zI)^{-1},
 \end{array}
 \end{equation}
 where $Y$ runs contractions in $\mathcal{N}$ satisfying inequalities $-G_{1}\leq Y\leq G_{2}$.
 \end{prop}
 \section{Application to some differential operators}
 Let us consider the multiplication operator $A$ in $\mathbb{L}_{2}(\mathbf{R}_{n})$
 by the continuous function $\varphi(k), \, k^{2}=k_{1}^{2}+...+k_{n}^{2},$ such that $\varphi(k)>0$ almost everywhere and
 \begin{equation}\label{cond1}
 \int\limits_{0}^{\infty} \frac{1}{(1+\varphi(k))^{2}}k^{n-1}dk<\infty .
 \end{equation}
 $A$ is a non-negative self-adjoint operator,
 \[
 \mathcal{D}(A)=\left\{f:\, \int_{\mathbb{R}_{n}}|1+\varphi (k)|^{2}|f(\mathbf{k}|^{2}d\mathbf{k}<\infty, \; f \in \mathbb{L}_{2}(\mathbf{R}_{n}) \right \}.
 \]
 In the sequel $\hat{\delta}$ stands for the unbounded linear functional in $\mathbb{L}_{2}(\mathbf{R}_{n})$, formally defined as follows:
 \[
 \hat{\delta}(f)=\int_{\mathbb{R}_{n}}f(\mathbf{k})d\mathbf{k}.
 \]
 Note that the domain of $\hat{\delta}$ contains $\mathcal{D}(A)$. Let us denote by $A_{0}$ the restriction of $A$ onto linear set
 \begin{equation}\label{domain}
 \mathcal{D}_{0}(A):=\left \{f: f\in \mathcal{D}(A), \;  \hat{\delta}(f)=0.\right \}
 \end{equation}
 The closure of $A_{0}\neq A$ and
 \[ \mathcal{N}= \left( \mathbb{L}_{2}(\mathbf{R}_{n}) \ominus (I+A)\mathcal{D}_{0}(A) \right) = \left \{ \xi \cdot \frac {1} {1+\varphi(k)}, \; \xi \in \mathbf{C} \right \} .
 \]
 Applying Theorem \ref{singular} yields
 \begin {prop}\label{multipl}
 $A$ is the unique non-negative self-adjoint extension of $A_{0}$ that is $A$ has no non-negative singular perturbations if and only if
 \begin{equation} \label{diverg}
 \begin{array}{ccc}
   \int\limits_{0}^{\infty} \frac{1}{\varphi(k)(1+\varphi(k))}k^{n-1}dk=\infty  & \mathrm{and} & \int\limits_{0}^{\infty} \frac{1}{(1+\varphi(k))}k^{n-1}dk=\infty.
 \end{array}
 \end{equation}
 \end{prop}
 Put $\varphi (k)=k^{2}$ and let $n=2$. Then the both integrals in Proposition \ref{multipl} are divergent. Hence the restriction $A_{0}$ of the operator $A$ of multiplication by $k^{2}$ in $\mathbb{L}_{2}(\mathbf{R}_{2})$ onto the linear set (\ref{domain} has unique non-negative self-adjoint extension  in $\mathbb{L}_{2}$. Note that the multiplication operator by  $k^{2}$ in $\mathbb{L}_{2}(\mathbf{R}_{n}$ is isomorphic to the self-adjoint Laplace operator $-\Delta$ in $\mathbb{L}_{2}(\mathbf{R}_{n}$ and its concerned here restriction $A_{0}$ is isomorphic to the restriction $-\Delta$ onto the Sobolev subspace $\mathbb{H}_{2}^{2}\left( \mathbf{R}_{n}\setminus\{0\}\right)$. As follows, the self-adjoint Laplace operator in $\mathbb{L}_{2}(\mathbf{R}_{2}$ has no non-negative singular perturbations with support at one point of $\mathbf{R}_{2}$.

 However, the non-negative singular perturbations of $-\Delta$ in $\mathbb{L}_{2}(\mathbf{R}_{2}$ with support at two or more points do already exist. For example, let us consider there the restriction $A_{0}$ of the multiplication operator by $k^{2}$, for which the defect subspace $\mathcal{N}$ is one-dimensional and consists of functions collinear to
 \[
 e_{0}(\mathbf{k})=\frac{1-\exp(-i(\mathbf{k}\cdot\mathbf{x}_{0}))}{1+k^{2}}, \; \mathbf{x}_{0} \in \mathbf{R}_{2}.
 \]
 In this case
 \[
  \begin{array}{c}
   \|e_{0}\|^{2}=\int_{\mathbf{R}_{2}}\frac{4\sin^{2}\frac{1}{2}(\mathbf{k}\cdot\mathbf{x}_{0})}{(1+k^{2})^{2}}\cdot d\mathbf{k}<\infty, \\  \left((I+A)A^{-1}e_{0},e_{0}\right)=\int_{\mathbf{R}_{2}}\frac{4\sin^{2}\frac{1}{2}(\mathbf{k}\cdot\mathbf{x}_{0})}{k^{2}(1+k^{2})}\cdot d\mathbf{k}<\infty, \\
   \left((I+A)e_{0},e_{0}\right)=\int_{\mathbf{R}_{2}}\frac{4\sin^{2}\frac{1}{2}(\mathbf{k}\cdot\mathbf{x}_{0})}{1+k^{2}}\cdot d\mathbf{k}=\infty.
 \end{array}
 \]
 Hence $G_{1}=\|e_{0}\|^{2}\cdot\left((I+A)e_{0},e_{0}\right)^{-1}>0$, but $G_{2}=0$. As follows, the concerned restriction $A_{0}$ of the multiplication operator {A} by $k^{2}$ has non-negative self-adjoint extensions in $\mathbb{L}_{2}(\mathbf{R}_{2}$ others then $A$ and $A$ is the maximal element in the set of these extensions. It remains to note that $A$ is isomorphic to the self-adjoint Laplace operator $-\Delta$ in $\mathbb{L}_{2}(\mathbf{R}_{2})$ and $A_{0}$ is isomorphic to the restriction of this $-\Delta$ on the subset of function $f(\mathbf{x})$ from $\mathcal{D}(-\Delta)$ satisfying conditions
\[\begin{array}{c}
    \underset{|\mathbf{x}|\rightarrow 0}{\lim}(\ln|\mathbf{x}|)^{-1}f(\mathbf{x})-\underset{|\mathbf{x}-\mathbf{x}_{0}|%
  \rightarrow 0}{\lim}(\ln|\mathbf{x}-\mathbf{x}_{0}|)^{-1}f(\mathbf{x})=0 ,  \\
    \underset{|\mathbf{x}|\rightarrow 0}{\lim}\left[f(\mathbf{x})-\ln|\mathbf{x}|\underset{|\mathbf{x'}|\rightarrow 0}{\lim}(\ln|\mathbf{x'}|)^{-1}f(\mathbf{x'})\right]- \\
    \underset{|\mathbf{x}-\mathbf{x}_{0}|%
  \rightarrow 0}{\lim}\left[f(\mathbf{x})- \ln|\mathbf{x}-\mathbf{x}_{0}|\underset{|\mathbf{x'}-\mathbf{x}_{0}|%
 \rightarrow 0}{\lim}(\ln|\mathbf{x'}-\mathbf{x}_{0}|)^{-1}f(\mathbf{x'})\right]=0.
  \end{array}
\]

Put now as above $\varphi (k)=k^{2}$ and let $n=3$. Then the first integral in Proposition \ref{multipl} is convergent while the second one as before divergent. Hence the restriction $A_{0}$ of the operator $A$ of multiplication by $k^{2}$ in $\mathbb{L}_{2}(\mathbf{R}_{3})$ onto the linear set (\ref{domain} has infinitely many non-negative self-adjoint extension  in $\mathbb{L}_{2}(\mathbf{R}_{3})$. As follows, the self-adjoint Laplace operator in $\mathbb{L}_{2}(\mathbf{R}_{3})$ has infinitely many non-negative singular perturbations with support at one point of $\mathbf{R}_{3}$ and the standardly defined Laplace the maximal element in the set of this perturbation.

As the next example we consider the multiplication operator $A$ by $k^{2l}$ in $\mathbb{L}_{2}(\mathbf{R}_{n})$ assuming that $4l\leq n+1$. $A$ is isomorphic to the polyharmonic operator $(-\Delta)^{l}$ in $\mathbb{L}_{2}(\mathbf{R}_{n})$. Let us consider the restriction $A_{0}$ of $A$ with the domain (\ref{domain}) that is non-negative symmetric operator which is isomorphic to the restriction of the polyharmonic operator $(-\Delta)^{l}$ onto the Sobolev subspace $\mathbb{H}_{2l}^{2}\left( \mathbf{R}_{n}\setminus\{0 \}\right)$. Applying Theorem \ref{singular} and Proposition \ref{multipl} results in the following proposition.
\begin{prop}
If $n<2l$ then there are infinitely many non-negative singular perturbations of $(-\Delta)^{l}$ associated with the one-point symmetric restriction $A_{0}$ and $(-\Delta)^{l}$ is the minimal element in the set of the non-negative extensions of $A_{0}$ in $\mathbb{H}_{2l}^{2}\left( \mathbf{R}_{n}\setminus\{0\}\right)$.

If $n=2l$ then $(-\Delta)^{l}$ has no such perturbations in $\mathbb{H}_{2l}^{2}\left( \mathbf{R}_{n}\setminus\{0\}\right)$ .

If $n>2l$ then there is the infinite set of non-negative singular perturbations of $(-\Delta)^{l}$ associated with  $A_{0}$ and  for those  as non-negative extensions of $A_{0}$  in the set of the  in $\mathbb{H}_{2l}^{2}\left( \mathbf{R}_{n}\setminus\{0\}\right)$ the operator $(-\Delta)^{l}$ is the maximal element.
\end{prop}
\newpage
\bibliographystyle{amsplain}

\end{document}